\documentclass{ws-ijmpe}

\begin{document}

\markboth{Ionel Stetcu, Bruce R. Barrett,
Petr Navr\'atil, and Calvin W. Johnson}{Electromagnetic transitions with effective operators}
%
\catchline{}{}{}{}{}
%

\title{ELECTROMAGNETIC TRANSITIONS WITH EFFECTIVE OPERATORS}

\author{IONEL STETCU and BRUCE R. BARRETT}

\address{Department of Physics, University of Arizona, \\
P.O. Box 210081, Tucson, Arizona 85721, U.S.A.}

\author{PETR NAVR\'ATIL}
\address{University of California, Lawrence Livermore National Laboratory,\\
Livermore, California 94551, U.S.A.}
\author{CALVIN W. JOHNSON}
\address{Department of Physics, San Diego State University, San Diego, California 92182, U.S.A.}


\maketitle

\begin{history}
\received{(received date)}
\revised{(revised date)}
\end{history}

\begin{abstract}
In the no-core shell model formalism we compute effective one- and
two-body operators, using the Lee-Suzuki procedure within the two-body cluster approximation. We evaluate the validity of the latter through calculations in reduced model spaces. In particular, we test the results for the two-body system and find that indeed the effective operators in the reduced space reproduce the expectation values or transition strengths computed in the full space. On the other hand, the renormalization for operators in the case of $^6$Li is very weak, suggesting the need for higher-body clusters in computing the effective interaction.
\end{abstract}

\section{Introduction}

One of the important steps in validating any theoretical model is a good description of the experimental data. This has to include, aside from the energy spectrum, observables and transition strengths, which provide an important test of the theoretical wave functions.

Starting from first principles, the no-core shell model (NCSM) has been successful in reproducing with very good accuracy the observed energy spectrum in light nuclei. Thus, within the NCSM framework, one starts with a realistic nucleon-nucleon (NN) interaction,\cite{c12} more recently adding also three-nucleon forces,\cite{marsden2002,navratil2003} and, by employing a similarity transformation, briefly reviewed in Sec. \ref{formalism}, obtains an effective interaction in a restricted model space. The initially infinite dimensional problem then becomes numerically tractable, so that diagonalization in a large, translationally invariant basis provides the theoretical spectrum as well as the wave functions of ground and excited states. If the transformation is exact, the energy spectrum and expectation values of other observables are preserved.

Much effort has been directed toward computing effective interactions, and with increasing computing power and more efficient shell-model codes, we are able nowadays to describe heavier and heavier systems in larger and larger model spaces. Less progress has been made, however, in the direction of consistently computing other effective operators within the NCSM. The only operators thus computed have been the effective point-proton radius,\cite{c12} relative kinetic energy,\cite{benchmark} and the NN pair density.\cite{benchmark,npn} However, important operators such as semi-leptonic weak and electromagnetic have not been consistently treated in the NCSM, and this is a possible explanation for the failure to describe adequately transition strengths.\cite{c12}

Historically, the method to obtain agreement with the experimental transition strengths was to use bare operators, with enhanced (e.g., $E2$) or quenched (e.g., Gamow-Teller) effective charges or couplings instead of the bare ones. But perturbation theory fails to predict the needed phenomenological charges;\cite{osnes} considerable success was however reported within the NCSM framework in a restricted model, where the effective quadrupole charges have been shown to have the expected values in a single harmonic oscillator shell.\cite{navratil1997} In the present paper, we test an implementation of general effective operators within the NCSM, by performing calculations in two- and many-body systems in limited model spaces.

\section{Formalism}
\label{formalism}

In the NCSM, all nucleons are treated on an equal footing. In this work, 
we consider two-body interactions only, that is, the intrinsic Hamiltonian describing the many-body system is 
\begin{equation}
H_{A}=\frac{1}{A}\sum_{i>j}\frac{(\vec p_i-\vec p_j)^2}{2m}+\sum_{i>j}V^{NN}_{ij},
\label{intrHam}
\end{equation}
where $m$ is the nucleon mass, and $V^{NN}_{ij}$ the bare NN interaction, such as the Argonne potentials in coordinate space\cite{argonne} or the non-local CD-Bonn.\cite{bonn}

Adding a center-of-mass (CM) Hamiltonian to the $A$-body Hamiltonian (\ref{intrHam}) does not change the intrinsic properties of the system. We bind the CM through a harmonic oscillator (HO), so that the total Hamiltonian can be cast in the form
\begin{eqnarray}
\lefteqn{
H_A^\Omega=H_A+\frac{\vec P^2}{2mA}+\frac{1}{2}mA\Omega^2R^2 \nonumber}\\
& &=\sum_{i=1}^A \left[ \frac{\vec{p}_i^2}{2m}
+\frac{1}{2}m\Omega^2 \vec{r}^2_i
\right] + \sum_{i<j=1}^A \left[ V^{NN}_{ij}
-\frac{m\Omega^2}{2A}
(\vec{r}_i-\vec{r}_j)^2
\right] \; .
\label{intrCM}
\end{eqnarray}
This introduces a pseudo-dependence on the HO frequency $\Omega$; however, the cluster approximation described below will introduce a real dependence on $\Omega$.

Following Da Providencia and Shakin\cite{PS64} and Lee,
Suzuki and Okamoto,\cite{LS80} we construct a similarity transformation able to accommodate the short-range two-body correlations by introducing an antihermitian operator $S$, so that the effective Hamiltonian ${\cal H}$ is given by
\begin{equation}
{\cal H}=e^{-S}H_A^\Omega e^S.
\end{equation}
In principle, both $S$ and ${\cal H}$ are $A$-body operators (even if originally we started with only two-body interactions), but determining $S$ exactly would be as hard as solving the original problem. We will return to the calculation of $S$ below; for now we assume that no approximation has been made.

The infinite Hilbert space associated with the system can be split into the finite model, or $P$-space, and the complementary, or $Q$-space, with the projectors $P$ and $Q$ spanning the entire space, $P+Q=1$. We determine the previous transformation by requiring that the new Hamiltonian completely decouples the $P$ and $Q$ spaces,
\begin{equation}
Q {\cal H} P=0,
\end{equation}
in addition to the conditions\cite{UMOA} $P S P=QSQ=0$. With these restrictions, the operator $S$ can be formally written\cite{UMOA} by means of another operator $\omega$ as
\begin{equation}
S=\mathrm{arctanh} (\omega-\omega^\dagger),
\end{equation}
with $Q\omega P=\omega$. In terms of the new operator $\omega$, the effective Hamiltonian in the $P$ space becomes
\begin{equation}
H_{eff}=P{\cal H}P=\frac{P +P\omega^\dagger Q}{\sqrt{P+\omega^\dagger\omega}}
H_A^{\Omega}\frac{P+Q\omega P}{\sqrt{P+\omega^\dagger\omega}},
\label{effHam}
\end{equation}
and, analogously, any arbitrary operator can be written in the $P$ space as\cite{navratil1993}
\begin{equation}
O_{eff}=P{\cal O}P=\frac{P +P\omega^\dagger Q}{\sqrt{P+\omega^\dagger\omega}}
O\frac{P+Q\omega P}{\sqrt{P+\omega^\dagger\omega}}\;.
\label{effOp}
\end{equation}

In order to determine the effective operators, one has to compute the transformation operator $\omega$, which is equivalent to determining $S$. But before addressing this problem, we would like to comment on the last two equations. We mentioned in the introduction that very little progress has been made regarding effective operators, while one can compute with very good accuracy the effective interactions. While it is true that Eqs. (\ref{effHam}) and (\ref{effOp}) are formally similar, the difference is that the interaction is given in relative coordinates, while a general operator is usually defined in the single-particle basis. For the effective interaction (\ref{effHam}), the problem can be broken down and solved one $({\cal J},T)$ channel at a time, the dimensions involved being much smaller than for the initial problem. The same procedure cannot be applied, however, for a general operator in the single-particle basis. The solution is to transform $\omega$ to the single-particle basis, but this involves a complicated approach which includes a large number of states. The problem simplifies again for special operators which are given in relative coordinates, such as the point-proton radius, relative kinetic energy, or NN pair density, and these have been investigated,\cite{c12,benchmark,npn} as previously noted.
Finally, note that because of Eq. (\ref{effOp}), a one-body operator will end up having non-zero genuine two-body elements in the model space.

We return now to the problem of determining $\omega$. Note that this operator connects vectors in $P$ to vectors in the $Q$ space. Therefore, a simple way to compute $\omega$ is\cite{c12}
\begin{equation}
\langle \alpha_Q|\omega|\alpha_P\rangle =\sum_{k}\langle \alpha_Q|k\rangle
\langle \tilde k|\alpha_P\rangle,
\label{omega}
\end{equation}
where $|\alpha_P\rangle$ and $|\alpha_Q\rangle$ are the basis states of the $P$ and $Q$ spaces, respectively, the $|k\rangle$ are the eigenvectors of the Hamiltonian
in the full space $H_A^\Omega|k\rangle=E_k|k\rangle$, and $\langle\alpha_P|\tilde k\rangle$ is the matrix element of the inverse overlap matrix $\langle\alpha_P|k\rangle$, that is $\sum_{\alpha_P}\langle k'|\alpha_P\rangle\langle \alpha_P|\tilde k\rangle=\delta_{kk'}$.

Note that computing $\omega$ by means of Eq. (\ref{omega}) requires knowledge of the solutions to the initial $A$-body problem, that is the eigevectors $|k\rangle$, which are actually our goal. While formally everything was exact up to now, in order to use this procedure, we have to make one approximation. That is, instead of using the eigenvectors $|k\rangle$ for the $A$-body problem, we solve these for the $a$-body system, with $a<A$. This is called the cluster approximation, and in the limit $a\to A$, the solution becomes exact. We point out that the cluster approximation induces a real dependence of the effective interaction upon the center of mass HO frequency; the solution to this problem is to search for a range of $\Omega$ values over which the results vary only slightly, i.e., are weakly $\Omega$ dependent.

The interaction obtained by means of Eq. (\ref{effHam}) at the $a$-body level is then diagonalized by means of the Lanczos algorithm in the many-body space spanned by an anti-symmetric, translationally invariant basis of $A$ particles. As noted, if $a=A$, one obtains the exact solutions. The way to reach convergence to the full space solution for $a<A$ is by increasing the dimension of the model space $P$, keeping the cluster approximation at the same level. (Note that when $P$ approaches the full space, the effective interaction approaches the bare one.) Although results calculated at the three-body cluster level have been recently reported,\cite{navratil2003,threecl} in this work we restrict ourselves to the two-body cluster, or in other words, $a=2$. (Three-body cluster calculations are computationally very demanding for obtaining the nuclear spectra and much more so in the case of other operators.)

Finally, when computing the effective operators by means of (\ref{effOp}), we have to make one additional approximation. That is because, as mentioned before, for arbitrary operators one cannot work in the relative system, and one needs to go to two-body matrix elements described in the single particle basis. This involves a large number of two-body matrix elements from the $Q$ space, and even for the simplest case one cannot take all of them into account. The solution is to include a restricted number of basis states from the $Q$ space, that is one HO shell at a time, and observe the convergence in the values of interest. For example, for a $2 \hbar \Omega$ $P$-space calculation, we include only states from $4\hbar\Omega$, $6\hbar\Omega$,etc. In each case we compute the two-body matrix elements of the effective operator, and when we do not see a change in their values, we assume that we have reached convergence.

\section{Results and discussion}

We apply first the procedure described above to the two-body system. In this case, the two-body cluster is the exact transformation, obtaining in the restricted space the exact expectation values or transition strengths as in the full space.

\begin{figure}
\centering\includegraphics*[scale=0.5]{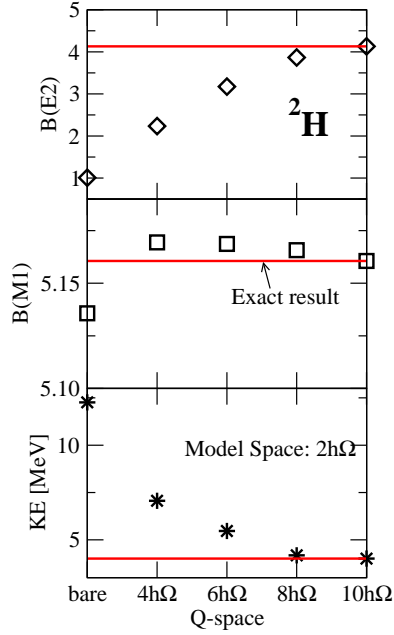}
\caption{Transition strengths of effective $E2$ and $M1$ operators, and the ground-state expectation value of the effective kinetic energy: evolution with the size of the $Q$-space included. The left-most values were obtained utilizing bare operators in the $2\hbar\Omega$ model space, while the straight lines mark the exact value obtained in the full space.
\label{H2r}} 
\end{figure}

Note that $\omega$ is a two-body operator, and, therefore, in order to compute effective one-body operators, we have to write the latter in a particle-number dependent two-body form:
\begin{equation}
O^{(1)}=\sum_{ab}\langle a | O^{(1)}|b\rangle a_a^\dagger a_b =\frac{1}{A-1}\sum_{{ab}{cd}}\langle a | O^{(1)}|b\rangle \delta_{cd} a_a^\dagger a_c^\dagger a_d a_b,
\label{ob}
\end{equation}
with the appropriate anti-symmetrization of the two-body matrix elements. In Eq. (\ref{ob}), the indexes of summation run over all single particle states, $a^\dagger$ and $a$ are the usual fermion creation and annihilation operators, respectively, and $\delta$ is the Kronecker symbol. 

In the two-body system, the two-body cluster approximation is exact; consequently, we have to regain in the restricted space exactly the values in the full space. For testing purposes, we assume that the full space is  the $10\hbar\omega$ shell-model space, in which we use the bare, isospin-independent Argonne V8' interaction.\cite{argonne} We emphasize that although we specifically employ one proton and one neutron in our calculations, this space does not give a realistic description of the deuteron, and is for testing purposes only. We then transform to a $2\hbar\omega$ model space, computing the effective interaction. Following the procedure described in the previous section, we compute $E2$ and $M1$ transition operators, as well as the effective relative kinetic energy by including one shell at a time from the $Q$ space. Figure \ref{H2r} shows that indeed when one includes most of the states in the $Q$ space, the effective operators reproduce the values in the full space. One might argue that the convergence is very slow, and that one needs most of the states to reach convergence. This is, however, most likely an artifact of the smallness of the full space, which makes the contribution of most $Q$-states equally important.

In many-body systems, the situation is different from the two-body system because the transformation from the full to the model space is not exact. We investigate the influence of the two-body cluster approximation in $^6$Li, by assuming that the full space is $8\hbar\omega$ and the interaction has a Gaussian form (due to the short-range repulsive core of realistic bare potentials, we cannot perform meaningful calculations in such a small space). We then compute the effective interaction and operators in $2\hbar\Omega$ and $4\hbar\Omega$ model spaces, and summarize the results in Table \ref{table:li6}. Note that because the initial full space is rather small, we have a more important contribution from higher order clusters. We mention however that although the ground state energy for the $4\hbar\Omega$ model space still differs by a few percents from the full space ($8\hbar\Omega$), the spacing between the states is about the same. All shell-model calculations have been performed with a descendant of the MFD code.\cite{MFD}

\begin{table}
\caption{\label{table:li6} $^6$Li: $B$ values of select $E2$ and $M1$ transitions, and expectation values of bare and effective relative kinetic operators. For comparison, we list the corresponding results with bare operators in the full space.}
\begin{tabular}{|c|cc|cc|c|}
\hline
Model Space & \multicolumn{2}{c|}{$2\hbar\Omega$} & \multicolumn{2}{c|}{$4\hbar\Omega$}&
Full Space \\
\cline{2-3} \cline{4-5}
 & Bare Op. & Eff. Op. & Bare Op. & Eff. Op. & ($8\hbar\Omega$)\\
 \hline
$B(E2; 1_1^+0\rightarrow 1_1^+0)$ &
 0.0080 & 0.0079 & 0.0097& 0.0097& 0.0364\\
$B(E2; 3_1^+0\rightarrow 1_1^+0)$ & 
 1.9512 & 1.9315 & 1.7575& 1.7548 & 1.5750\\ 
$B(E2; 1_2^+0\rightarrow 1_1^+0)$ &
 0.8855 & 0.8765 & 0.7130 & 0.7117& 0.5791\\
$B(E2; 2_1^+0\rightarrow 0_1^+1)$ &
 0.0870 & 0.0866 & 0.0865& 0.0867& 0.0820\\
\hline
$B(M1; 0_1^+1\rightarrow 1_1^+0)$ &
 5.3926 & 5.3549 & 5.4209& 5.4051& 5.3346\\
\hline
$\langle 1_1^+0 |T_{rel}| 1_1^+0\rangle$ &
 116.15 & 118.46 & 124.66 & 125.84 & 135.84\\
$\langle 0_1^+1 |T_{rel}| 0_1^+1\rangle$ &
 116.38 & 118.58 & 124.56 & 125.67 & 135.19\\
\hline
\end{tabular}
\end{table}

Table \ref{table:li6} shows insignificant differences between the results obtained with bare or effective operators (note that in this case we show only the values when we included the full $Q$ space). We have to point out that previous calculations have produced very strong renormalization of the kinetic energy operator in $^4$He\cite{benchmark} and $^{12}$C,\cite{unpublished} but here we find it to be very weak. The explanation lies again in the smallness of the full space which makes the higher-order clusters, most likely up to six particles, much more important than in realistic calculations. Nonetheless, we have unpublished results in realistic calculations in $^{12}$C for effective electromagnetic operators,\cite{unpublished} which show similar behavior as the results in Table \ref{table:li6}. This suggests that for such operators the two-body cluster does not produce enough renormalization, to be confirmed by further investigations in realistic model spaces, using realistic interactions. Note that the relative kinetic energy is a short range operator, while the $E2$ transition and point-proton radius (for which was also found a very weak renormalization) operators are long-range operators. It is, however, known from effective interaction calculations that the two-body clusters renormalize the short-range correlations of the NN interaction, so it follows that the renormalization for long-range operators should be weaker. What is somehow surprising is the extremely small renormalization of the $E2$ operator, which should have also a short-range component (there is no surprise that the $M1$ operator does not renormalize much, as it does not mix different shells, and, hence, the contribution from outside the model space should be very small).

\section{Conclusions}

We have implemented the NCSM formalism to general one- and two-body operators. For effective interactions we use a procedure which is equivalent with using the entire $Q$ space (for details we refer the interested reader to previous work\cite{c12}). Arbitrary operators present a significant difference, as, in general, one cannot use the whole complementary $Q$ space. Hence, we developed a convergence procedure by adding $Q$ states one shell at a time. While in the calculation presented the convergence appears to be slow, we expect that this is due to the limitation of the size of the model space, where higher-order clusters play an unrealistically big role.

In the two-body system, where there is no approximation in computing the transformation from the full to the model space, we regain in the exact $B(El)$, $B(Ml)$, or expectation values by employing effective operators. Furthermore, these renormalized results differ significantly from the ones obtained by using only the bare operators.

Our test in a many-body system, where we employ only the two-body cluster approximation, shows a weak renormalization for electromagnetic operators as well as for the relative kinetic energy. The latter has been previously computed in realistic systems, and the renormalization was found to be significant. We explain the present results by the size of the full space, which is very small in our example, making the higher-order cluster contributions more important than in realistic situations. However, for electromagnetic transition operators, we found similar results in a realistic calculation for $^{12}$C,\cite{unpublished} which would suggest that, compared with the effective interaction, the higher-order clusters play a bigger role in computing long-range effective operators. This is a hypothesis remaining to be verified in realistic calculations. As a general caveat, any truncation of the space could induce 
effective operators with non-negligible higher-body correlations. Consequently, the renormalization properties of each operator, especially those with little or no experimental data available (e.g., double-$\beta$ decays), need to be studied separately.

Nevertheless, the theory of effective operators is important for at least two sets of next generation experiments. The first is the investigation of nuclei far-from-stability through the projected rare isotope accelerator (RIA), with important applications to nucleosynthesis. The second is to double-$\beta$ decay experiments, which will provide data on the mass of the lightest neutrino.\cite{2beta} This motivates us even more to pursue this issue further, by investigating, for example, semi-leptonic operators at finite momentum transfer  which should have short-range components, renormalizable at the two-body cluster level.

\section*{Acknowledgments}

I.S. and B.R.B acknowledge partial support by NFS grants PHY0070858 and PHY0244389.
This work was performed in part under the auspices of the U. S. Department of Energy by the University of California, Lawrence Livermore National Laboratory under contract No. W-7405-Eng-48. P.N. received support from LDRD contract 04-ERD-058. C.W.J. acknowledges support by USDOE grant No. DE-FG02-96ER40985.

\end{document}